# Algorithmic complexity in computational biology: basics, challenges and limitations

Davide Cirillo[#,1,*], Miguel Ponce-de-Leon[#,1], Alfonso Valencia[1,2]

[1] *Barcelona Supercomputing Center (BSC), C/ Jordi Girona 29, 08034, Barcelona, Spain*
[2] *ICREA, Pg. Lluís Companys 23, 08010, Barcelona, Spain.*

# Contributed equally
* corresponding author: davide.cirillo@bsc.es (Tel: +34 934137971)

*Key points*:

- Computational biologists and bioinformaticians are challenged with a range of complex algorithmic problems.
- The significance and implications of the complexity of the algorithms commonly used in computational biology is not always well understood by users and developers.
- A better understanding of complexity in computational biology algorithms can help in the implementation of efficient solutions, as well as in the design of the concomitant high-performance computing and heuristic requirements.

*Description of the author(s):*

**Davide Cirillo** is a postdoctoral researcher at the Computational biology Group within the Life Sciences Department at Barcelona Supercomputing Center (BSC). Davide Cirillo received the MSc degree in Pharmaceutical Biotechnology from University of Rome 'La Sapienza', Italy, in 2011, and the PhD degree in biomedicine from Universitat Pompeu Fabra (UPF) and Center for Genomic Regulation (CRG) in Barcelona, Spain, in 2016. His current research interests include Machine Learning, Artificial Intelligence, and Precision medicine.

**Miguel Ponce de León** is a postdoctoral researcher at the Computational biology Group within the Life Sciences Department at Barcelona Supercomputing Center (BSC). Miguel Ponce de León received the MSc degree in bioinformatics from University UdelaR of Uruguay, in 2011, and the PhD degree in Biochemistry and biomedicine from University Complutense de, Madrid (UCM), Spain, in 2017. His current research interests include Systems Biology, Computational Modeling and Simulations, and Precision medicine.

**Alfonso Valencia** is the Director of Life Sciences Department and Computational biology Group Leader at Barcelona Supercomputing Center (BSC). Alfonso Valencia is an ICREA Professor and a senior scientist with activity in various areas of bioinformatics and computational biology. He is the founder of the BioCreative text-mining challenge, and of the International Society for Computational biology (ISCB). He is a member of EMBO, and of the Scientific Advisory Board (SAB) of the Swiss Institute of bioinformatics, among others. He is the Director of the Spanish Institute for bioinformatics, the Spanish node of the European bioinformatics infrastructures (ELIXIR). He is co-Executive Editor of the journal bioinformatics, Oxford University Press.

## Abstract

Computational problems can be classified according to their algorithmic complexity, which is defined based on how the resources needed to solve the problem, e.g. the execution time, scale with the problem size. Many problems in computational biology are computationally infeasible in the sense that the exhaustive search for the optimal solution is prohibitive in practical terms. As a consequence, these problems are tackled through heuristics and approximations aiming to overcome the exceeding computational requirements at the cost of providing suboptimal solutions. The importance of defining the computational complexity of computational biology algorithms is a topic rarely surveyed for broad audiences of bioinformaticians and users of bioinformatics tools. However, recognizing the underlying complexity of any algorithm is essential for understanding their potential and limitations. Thus, the aim of this review is to survey the main algorithmic solutions to intractable problems in computational biology, highlighting the importance of High-Performance Computing in this area.

## Introduction

With unprecedented advancements in high-throughput technologies and increased availability of vast amounts of data, computational biology and bioinformatics are facing the grand challenge of handling effective ways to process and analyze biomedical information on a large scale. A plethora of bioinformatics methods are ordinarily deployed, and their impact quantified [1]. For instance, comprehensive biomedical tool platforms are offered by the international research infrastructures ELIXIR (https://elixir-europe.org/platforms/tools) and Big Data to Knowledge (BD2K) (https://commonfund.nih.gov/bd2k/resources). Such a wide variety of available computational tools, often dedicated to single and very specific problems could be bewildering for the general users and, occasionally, developers. Long-standing problems, such as multiple sequence alignment (MSA), still defy existing computational solutions which are constantly improved through new implementations, now able to align millions of sequences on standard workstations [2]. Is this the final solution? Is the problem solved? Why not provide just the optimal solution to a given problem? The answers to these questions are far from being of little significance and lay at the heart of computational complexity theory, a branch of computer science concerned with the classification of computational problems according to their inherent difficulty. Despite its importance, general users, and frequently computational biologists and bioinformaticians, are often not familiar with computational complexity and related concepts.

The possibility to find an exact solution to a problem, as opposed to a suboptimal solution, largely depends on the nature of the problem itself, the specific instance of the problem to be solved, and the computational resources available. In this regard, some problems can be optimally solved in a reasonable time, while others are so complex that only suboptimal solutions can be achieved, generally through a number of possible practical approaches, known as heuristics. For instance, a simple problem is checking if a graph is connected, i.e. if any node can be reached from any other node by traversing the graph. Conversely, a complex problem is to find a path in a graph that visits each node exactly once, also known as the Hamiltonian path problem (**Figure 1A-B**). In the context of computational biology, the problem of assembling a genome from short overlapping fragments of DNA can be expressed as a Hamiltonian path problem. This algorithm, known as the Overlap-layout-consensus, constructs a graph where DNA fragments correspond to nodes and a directed edge is set between two nodes if there is prefix-suffix overlap between corresponding DNA fragments (**Figure 1C-D**). In the last step fragments are assembled by finding a path that traverses each node exactly once, i.e. a Hamiltonian path (**Figure 1E**). Other approaches for de-novo genome assembly using different graph representations, such as the Bruijn graphs, have also been extensively used [3,4].

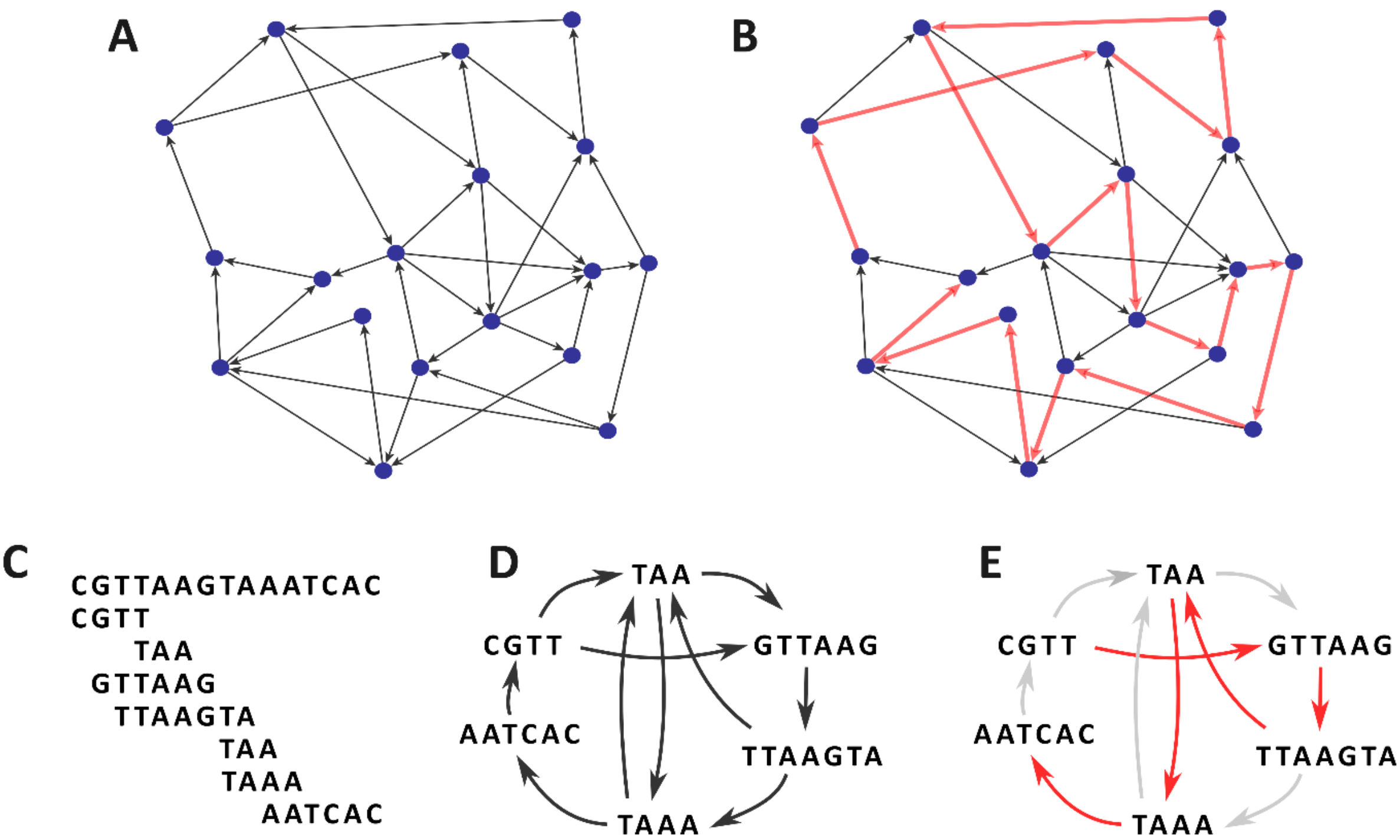

***Figure 1****. The Hamiltonian path problem. (A) The problem consists of identifying the path that visits all the other nodes exactly once. (B) A solution to the problem is reported. (C) and (D) show a toy example of genome assembly formulated using the Overlap consensus layout algorithm that relies on solving the Hamiltonian path problem using fragments overlaps. One possible solution is reported in (E).*

The mathematical puzzle of the Hamiltonian path can be further complicated by adding distances between the nodes and reformulated to minimize the total travelled distance. This classic graph problem is known as the Travelling Salesman Problem (TSP). Many problems in computational biology can be formulated as a TSP problem. For example, phylogenetic tree reconstruction consists in finding the minimum genetic distances for a given set of sequences, which can be expressed as a problem of finding the shortest cycle path that visits each node only once. In particular, given a set of aligned DNA sequences represented as vertices and connected if there is one mismatch (a.k.a, a DNA grid graph), the minimum amount of sequence changes corresponds to the minimum spanning tree, an approximated solution to the TSP [5]. Besides implicit phylogenetic models such as maximum parsimony, methods to estimate a phylogenetic tree can imply explicit models of DNA evolution.

Novel parallel architectures are being proposed to deal with the growth in computational complexity, with High-Performance Computing (HPC) being one of the most successful. In addition to these improvements in HPC, other computing paradigms such as DNA-based molecular computing [6] and quantum computing [7] are also active and promising areas of research enabling to greatly increase the efficiency of solving complex problems.

As explained in the next sections, the vast majority of bioinformatics problems are computationally complex, and therefore users and developers of bioinformatics algorithms would benefit from being acquainted with the concept of computational complexity. We review here what computational complexity is, and how it maps to the fundamental problems in computational biology, examining the main heuristic approaches, and the relation with HPC.

## Algorithmic complexity in computational biology

The efficiency of an algorithm, referred to as *computational complexity*, is the amount of resources required for its execution, these resources being the time needed to execute the algorithm (i.e. the number of elementary operations to perform) and space (i.e. the size of the required memory). The study of algorithmic complexity in computational biology provides directions for the implementation of efficient programs for processing, modelling, and analysing biological data.

Algorithmic complexity is expressed using the so-called asymptotic or “Big-O” notation. This notation expresses how the execution time of an algorithm grows as a function of the problem size $n$. It is worth noting that the asymptotic notation describes the behaviour of an algorithm in the worst-case scenario (i.e. using the most problematic instances of the problem) and when the size of the problem $n$ tends to infinity. The asymptotic notation defines distinct levels of algorithmic complexities, from low to high, namely constant time $O(1)$, logarithmic time $O(\log(n))$, linear time $O(n)$, quasilinear time ($O(n \log(n))$, quadratic time $O(n^2)$, exponential time $O(2^n)$, and factorial time $O(n!)$ complexity, as the most common functions (**Figure 2**). Notably, combinatorial problems solved through an exhaustive search, such as systematically visiting the nodes of a graph representing a phylogenetic tree, a metabolic pathway, or a protein interactome, have factorial time complexity [8].

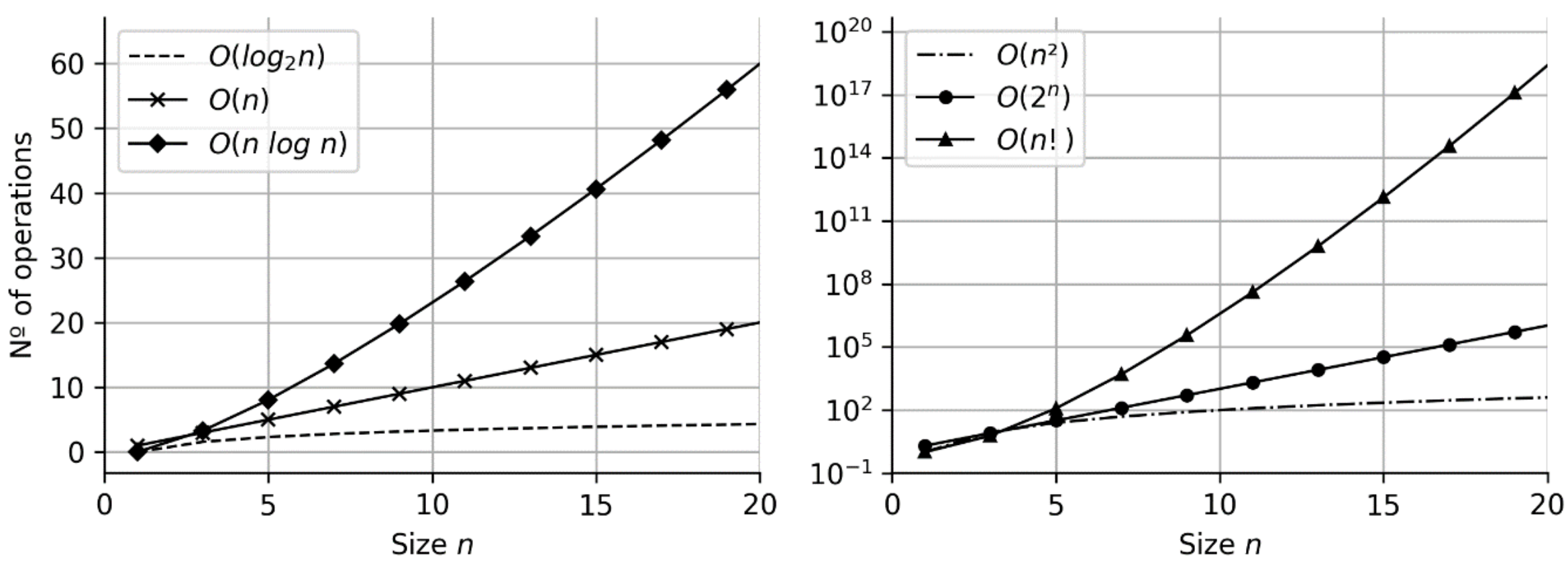

***Figure 2.*** *Common functions describing algorithmic complexity. The number of elementary operations is shown as a function of the number n of input elements (problem size), expressed using the asymptotic notation (O). On the left, low levels of algorithmic complexity, namely logarithmic time O(log(n)), linear time O(n), quasilinear time (O(n log(n)). On the right, high level of algorithmic complexity, namely quadratic time* $O(n^2)$*, exponential time* $O(2^n)$*, and factorial time O(n!).*

In general, only those algorithms that exhibit a polynomial time complexity, that is $O(n^k)$ for some non-negative integer k, are accepted to solve problems in a reasonable amount of time. Several polynomial algorithms have applications in computational biology, such as the illustrative examples reported in **Table 1**.

Despite the efforts to find an algorithm that solves any given problem in polynomial time, this pursuit is limited by the actual inherent complexity of the problem itself, such as the case of many problems for which finding the exact solution necessarily requires exponential time (see examples in the next section).

***Table 1***. *Examples of notable polynomial algorithms and their application in bioinformatics.*

| Problem | Algorithm | Application in Computational Biology | Complexity | Reference |
|---|---|---|---|---|
| Discrete Fourier Transform of arbitrary composite size $n$ | Cooley–Tukey algorithm | *Prediction of DNA-Protein interaction* | $O(n\log(n))$ | [9] |
| Shortest paths in a graph of $\|E\|$ edges and $\|V\|$ vertices | Dijkstra's algorithm | *Pathway prediction in metabolic and signalling networks* | $O(\|E\|\log\|V\|)$ | [10] |
| Matching a pattern of length $n$ in a string of length $m$ | Knuth-Morris-Pratt algorithm | *Exact pattern matching in DNA sequences* | $O(m+n)$ | [11] |
| Phylogenetic tree reconstruction | Neighbour Joining | *Phylogenetic inferences* | $O(n^3)$ | [12] |
| Finding Eulerian paths in a graph with \|E\| edges | Hierholzer's algorithm | *De-novo genome assembly* | $O(\|E\|)$ | [13] |
| Finding the most likely sequence L of hidden states S | Viterbi algorithm | *Protein family domains and gene finding* | $O(L\|S\|^2)$ | [14] |

### Multiple sequence alignment: a prototypical computational problem

Multiple sequence alignment (MSA) is one of the most important problems in computational biology. A biologically accurate sequence alignment is the one that maximizes the underlying evolutionary or structural relationships among the sequences. MSA uses estimations of the likelihood of amino acids substitutions described in the widely used PAM [15] and BLOSUM [16] matrices.

MSAs are crucial for genomic and proteomic analysis, including genome assembly, protein structure prediction, and phylogenetic reconstruction. They are used for the identification of conserved regions in biological sequences of amino and nucleic acids, revealing structural, functional, and evolutionary information [17]. With the present-day explosion of sequencing data [18], accurate MSA techniques have become a fundamental prerequisite for large-scale sequence comparisons, in particular taking into account the ongoing transnational projects for large-scale genomic data generation and sharing [19].

In the case of pairwise sequence alignment, dynamic programming [20] guarantees the optimal alignment for a defined set of scores for matches, mismatches, and gaps. Nevertheless, in dynamic programming the number of comparisons increases exponentially as the number of sequences increases, namely $O(2^n L^n)$ for $n$ sequences of average length L. This makes the problem infeasible in practical terms even for a small number of sequences. As a consequence, diverse strategies, or heuristics, have been proposed to overcome the computational burden associated with the problem. For instance, the Divide-and-Conquer Alignment [21] recursively cuts the original sequences into sub-sequences until these are short enough to be aligned optimally and ultimately concatenated. Given $n$ sequences with longest length $L$, the time needed to compute all the combinations of cut points is $O(L^{n-1})$, while $O(n^2L^2)$ is the time needed to compute all alignments, resulting in an overall complexity of $O(n^2L^2+L^{n-1})$, which reduces the execution time compared to dynamic programming (**Figure 3**).

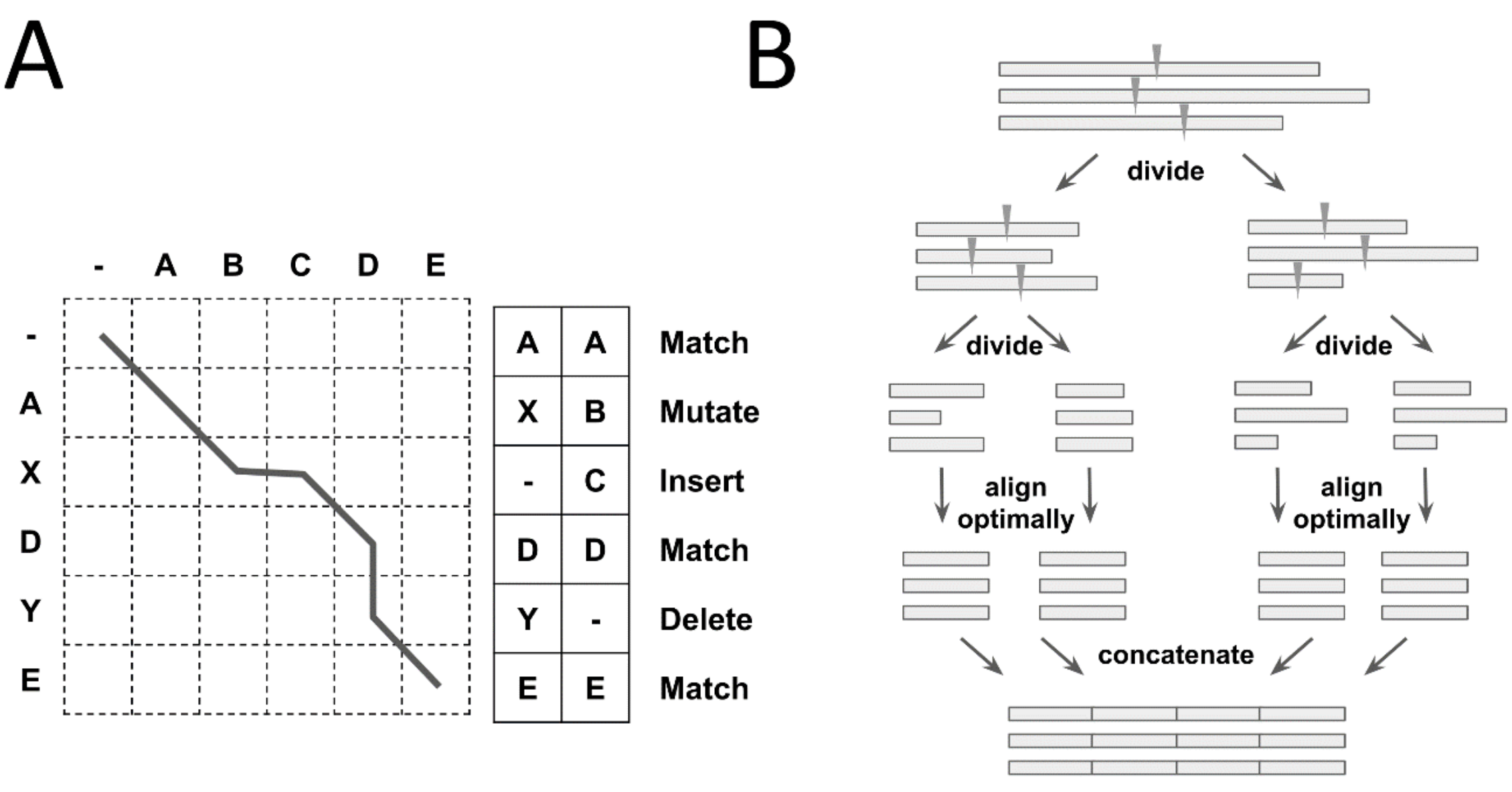

***Figure 3.*** *Schematic representation of (A) Dynamic Programming for pairwise sequence alignment, and (B) Divide-and-Conquer Alignment for MSA. Adapted from [22] and [23].*

The two examples of dynamic programming and Divide-and-Conquer Alignment, show how a possible solution to a given problem is not unique and strongly depends on the goals of the analysis. For instance, Divide-and-Conquer Alignment provides quality alignments at the cost of an execution time that scales exponentially with the number of sequences. Nevertheless, faster aligners rely on different algorithmic strategies, such as progressive alignment, which aligns the most similar sequences first, implemented in ClustalW, with complexity $O(n^2)$, and regressive alignment, which aligns the most dissimilar sequences first, implemented in the latest version of T-Coffee, with complexity $O(n)$ [2]. Of note, the case of progressive and regressive MSA is an illustrative example of two conceptually opposite approaches pursuing better performances for one single intrinsically difficult problem. Other algorithmic approaches to MSA are iterative alignment, such as Praline [24] and MUSCLE [25], with complexity $O(n^2L + nL^2)$; consistency-based alignment, such as MAFFT [26], with complexity $O(n \log n)$, and T-coffee [27], with complexity $O(n^3L)$; and structure-based alignment, such as Expresso [28], with complexity $O(n^3L)$.

**Complexity classes**

Computational problems can be divided into those for which an algorithm exists (decidable) and those for which an algorithm is proven to not exist (undecidable). Decidable problems, for which a yes/no answer is demanded, can be grouped into different classes according to their computational complexity.

A computational problem belongs to the complexity **class P** (Polynomial-time) if an algorithm exists for which the number of elementary operations needed to find the solution is bounded by a polynomial function of the problem input size (**Figure 4**). P problems are considered computationally tractable [32]. On the other hand, a problem belongs to the complexity **class NP** (Non-deterministic Polynomial-time) if an algorithm exists for which it can check in polynomial time whether a given solution is correct. Counterparts of NP problems are co-NP problems, for which excluding wrong solutions can be achieved in polynomial time.

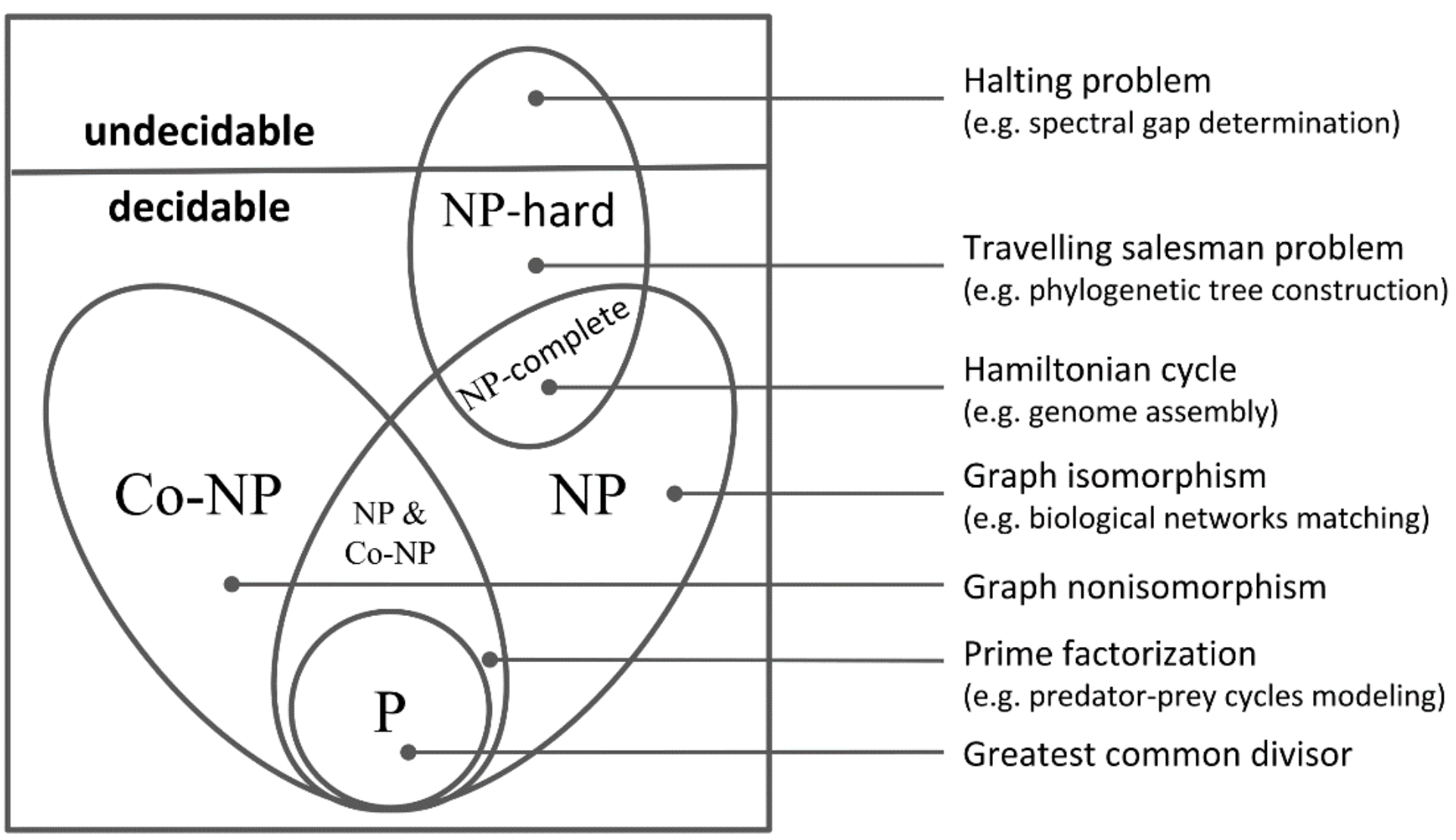

***Figure 4****. Decidability and complexity classes, under the unproven assumption that P≠NP [29]. General computational problems are reported with examples in computational biology. Algorithms with different levels of efficiency exist for problems with different levels of difficulty (P, NP, NP-complete, and NP-hard). For instance, prime factorization algorithms exist for the problem in systems biology of determining the emergence of prime-numbered cycles in predator-prey strategies [30]. Conversely, no algorithms exist for the problem in nanobiotechnology of determining the spectral gap in a growing atomic lattice, which corresponds to the algorithmically unsolvable Halting problem [31].*

The most difficult NP problems belong to the complexity **class NP-complete**, where the difficulty order is determined by the possibility of efficiently reducing one problem to another. Thus, if we could solve an NP-complete problem efficiently, we would be able to solve all NP problems efficiently. Problems that are at least as hard as NP-complete problems belong to the complexity **class NP-hard**. Hundreds of problems of practical relevance, especially in computational biology, fall into the NP-complete class [33,34]. The majority of these problems relate to graphs and mathematical programming, such as the Travelling Salesman Problem (TSP) presented in the Introduction. In the computational biology domain, different problems such as MSA or genome assembly can be related to TSP [35,36]

Examples of NP-complete problems in computational biology are MSA [37], phylogenetic tree reconstruction [38], and the general problem of finding global optima in nonlinear or combinatorial optimization [39], such as minimizing energy potentials (e.g. protein folding prediction) or large-scale parameter fitting (e.g. kinetic modelling in biochemical systems).

Other complexity classes include EXP problems, for which it takes exponential time or space to check the correct solution; PSPACE problems, which can be solved with an unlimited amount of time but using only a polynomial amount of space for memory; and BPP problems (or BQP for quantum computing), which can be solved probabilistically in polynomial time. In today's era of Big Data, the implementation of strategies for effective data compression and sustainable hardware solutions are fundamental to tackle problems related to space complexity and the amount of memory required to solve them [40,41].

### Complexity and machine learning

The widespread adoption of deep learning for computational biology problems [42] makes the complexity of machine learning algorithms an increasingly relevant and necessary issue in this area [43,44]. It has been recently demonstrated that determining whether a machine learning algorithm is able to make predictions about a large data set by sampling few data points (i.e. learnability) is an undecidable problem given the infinite ways of sampling the smaller set [45]. Despite the ability to model large volumes and a great variety of data, the application of deep learning to real-world problems, such as those arising in healthcare and medicine, poses several challenges in terms of computational resources. The ImageNet dataset, used for training Convolutional Neural Networks (CNNs) for image processing tasks, consists of more than $10^7$ images and over $10^4$ categories [46]. The Bidirectional Encoder Representations from Transformers (BERT) language model consists of 110M parameters [47]. As a consequence, deep learning is largely performed on distributed memory, with graphics processing units (GPUs) being the main hardware employed [48]. Recently, tensor processing units (TPUs), custom-designed chips developed by Google for deep learning applications and used for AlphaGo algorithm training [49] have considerably grown in popularity. These developments are paving the way to machine learning-specific hardware implementations, such as Edge TPUs for edge computing and Internet of Things (IoT), characterized by smaller size and low power consumption.

### Heuristics

In real-world problems, finding a sub-optimal solution is often the only option as it would take too much time to solve the problem exhaustively even if a polynomial algorithm is available. The design of approaches that trade optimality, accuracy and completeness for speed are generally known as heuristics. Table 2 summarizes different computational problems in computational biology as well as notable heuristic approaches used to find good-enough solutions.

***Table 2***. *Examples of heuristic approaches and methods for common complex algorithmic problems in computational biology.*

| Computational Biology problem | Heuristic approach | Reference |
|---|---|---|
| Genome assembly | Generation of a directed multigraph between neighbouring k-mers. | Eulerian paths [13] |
| Regulatory motif finding | Position frequency matrices inference using online expectation-maximization. | EXTREME [50] |
| Phylogeny reconstruction | Search of near-optimal parsimonious tree by iterative improvements. | Local search [51] |
| Structure prediction | Simulated annealing with structural constraints. | Rosetta [52] |
| Finding similarities in interaction networks | Subgraph estimates based on the sparse boundary frequencies. | Targeted Node Processing [53] |
| Discovering splicing variants | Splice junctions filtering based on minimum minor isoform frequency estimates. | TopHat [54] |
| Classification of metagenomic sequences | Asynchronous search and consensus-based on species, genus, and class taxonomy. | SMART [55] |
| Analysis of metabolic pathways | Stoichiometry and reaction rates constraints. | Constraint-based methods [56] |

Local search algorithms, as well as greedy and probabilistic techniques, are classical examples of heuristic approaches adopted to solve different problems [57]. For instance, the widely used agglomerative procedure in hierarchical clustering is an example of a greedy heuristic to the problem of clustering data points. On the other hand, approximation algorithms differ from other heuristics by the fact that they guarantee the quality of the solution by providing its distance to the optimal one. It is worth mentioning that approximation algorithms are only relevant to optimization problems, i.e. any problem where the best solution is found given an optimality criteria, usually defined by a cost or objective function to be maximized or minimized. In this way, it is possible to measure how far a solution is from the optimum, which is a factor known as the approximation ratio. Therefore, while approximation algorithms guarantee the quality of the returned suboptimal solution, more general heuristic approaches, such as local search algorithms, do not provide this information. An example of an approximation algorithm in the context of genomics is the Sorting by Reversals algorithm. This algorithm is used to find the smallest set of rearrangements (e.g., inversions and transpositions of fragments) between the genomes of two relative species [58].

As discussed above, finding an optimal MSA becomes computationally intractable as the number of sequences increases. As a consequence, all available algorithms for finding

MSA rely on some kind of heuristic [59]. For example, a polynomial-time approximation algorithm has been proposed, achieving alignments whose scores are at most 2-2/k times the optimum for k sequences [60]. This approximation ratio was further improved to 2-3/k [61] and 2-ℓ/k [62] for any fixed ℓ, before the problem was eventually proven to be NP-complete [37], meaning that the approximation ratio cannot be made arbitrarily close to 1. However, it was shown later that a polynomial-time approximation can be achieved if the alignment is guided by providing a binary phylogenetic tree [63]. These results show how important it is to achieve a more efficient and accurate heuristic for solving real problems of key importance in biomedical research.

Another classical example of a heuristic for solving NP-complete problems in computational biology is the Neighbour-joining algorithm, a distance-based method that iteratively chooses two nearest nodes as neighbours for the reconstruction of a phylogenetic tree [12]. Phylogenetic distances are generally computed from DNA sequences. While the general problem of reconstructing a phylogenetic tree for n taxa has proven to be NP-complete [5], the Neighbour-joining heuristic can find a good-enough tree with a polynomial complexity of $O(n^3)$.

The so called meta-heuristic methods [64], generally related to evolutionary computation and swarm intelligence, are commonly applied to solve nonlinear global optimization problems, such as parameter estimation in systems biology models [65]. Efficient meta-heuristic methods require extensive computational experiments in order to tune the heuristic parameters and the fitness function. A common approach to calibrate parameters relies on the application of the meta-heuristic to instances of the problem where the optimal solution is known beforehand. When the metaheuristic is used to find a solution to the test problem, its quality is compared to the exact solution. In this way, the performance of the meta-heuristic can be rigorously tested for different sets of parameters.

**High-performance computing in computational biology**

Establishing procedures that avoid the exponential explosion inherent in the exhaustive search approach is crucial to finding solutions to problems in computational biology. The employment of supercomputers to perform massively parallel computation ensures that the process will be completed in a reasonable amount of time, carrying out billions of operations per second. Such parallel computing on high-end hardware is generally referred to as High-Performance Computing (HPC). HPC resources can be particularly suitable to use heuristic strategies for solving NP problems of any kind as well as P problems with extremely large instances.

Increased transistor count and power dissipation advances made possible the rapid improvement in cost-performance of **sequential computing**, resulting in software algorithms traditionally constructed for serial computation [66]. Nonetheless, the power limit of a single chip rapidly proved insufficient to face innovations like multiple issues of instructions and dynamic execution, leading to the creation of the new architectural paradigm of **parallel computing**, in which many power-efficient processors, or cores, are placed on the same chip, or multicore microprocessor.

As parallel algorithms are more challenging to develop than sequential algorithms [66], several attempts to systematize parallel programming have been carried out. For instance, a number of common patterns of scientific computing, or computational motives, have been identified (**Table 3**) to be combined into complex parallel software systems [67]. As shown in Table 3, many different algorithms used in computational biology heavily rely on these motives. Programmers in high-performance computing look steadfastly towards Life Sciences applications in order to transfer the capabilities of parallel programming and explicit libraries such as MPI (www.mpi-forum.org/) and OpenMP (www.openmp.org/) and enhance innovation in health and biomedical research. The key challenge in this area is optimizing the synchronization of concurrent processes that determines different levels of parallelization for specific programs [81].

***Table 3.*** *Common computational motives [67] with examples of applications in computational biology suitable for implementation in a parallel computing environment.*

| Computational motives for parallel computing | Description | Computational biology applications |
|---|---|---|
| Dense linear algebra | Dense matrices or vectors | Linkage disequilibrium computation in genome-wide association studies [68] |
| Sparse linear algebra | Matrices or vectors in which most elements are zeroes | Stochastic chemical kinetics [69] |
| Spectral methods | Use combinations of basis functions to solve differential equations | Multiple sequence alignment [70] |
| N-body methods | Interaction between discrete points (particles) | Agent-based simulations of cellular behaviour [71] |
| Structured grids | Tessellation of n-dimensional Euclidean space using regular connectivity | Multi-dimensional organ simulations [72] |
| Unstructured grids | Tessellation of n-dimensional Euclidean space with irregular connectivity | Whole-body biomedical simulations [73] |
| Monte Carlo | Calculation based on repeated random trials (stochastic simulations) | Emission tomography image processing [74] |
| Combinational logic | Digital logic producing specified outputs from certain inputs | Allosteric receptor modelling [75] |
| Graph traversal | Visiting vertices in a graph | Biological network analysis [76] |
| Graphical models | Graph-representation of conditional dependencies | Biological network modelling [77] |
| Finite state machines | Interconnected set of states | Multicellular behaviour [78] |
| Dynamic programming | Recursively finding optimal solutions of the sub-problems of a larger problem | Multiple sequence alignment [79] |
| Backtrack and Branch-and-Bound | Regions of the search space with no interesting solutions are ruled out | RNA secondary structure prediction [80] |

Moreover, a great improvement in computational efficiency can be obtained by using vectorization in the implementation of an algorithm. Vectorization allows to write array operations (e.g. matrix or vector product) instead of using a loop that applies the same operation multiple times to different values. Although not always possible, when vectorization can be used to replace loops, the code will exploit the full potential of modern parallel architectures. These kinds of improvements are critical for getting better performance in specific problems (e.g. MSA and genome assembly) and also for scaling simulations of molecular and biological systems. For instance, molecular dynamics [82] or agent-based simulations of multicellular systems [83] are computationally demanding applications that usually need to numerically integrate systems of partial differential equations at each time iteration or solve other kinds of algebraic problems. Therefore, simulations of biological systems benefit from improving computational performance through the optimization of underlying numerical algorithms as well as from HPC-based implementations.

In recent years, several bioinformatics methods have greatly exploited parallel computing solutions, such as computational genomics tools for the identification of genes that contribute to phenotypic variation. Indeed, due to the increasing number of new variants emerging from the ever-increasing genome sequencing (e.g. single nucleotide polymorphism, insertions, deletions, copy number alterations), computing linkage disequilibrium (i.e., the non-random association of alleles at different loci) presents a major bottleneck in allele and haplotype frequencies calculation in large population counts. However, parallel implementations allow finding possible solutions in reasonable periods of time [68]. For instance, given the complexity and size of the new genomic data sets produced by single-cell sequencing [84], efficient algorithmic HPC implementations are going to become a requirement in this area.

## Discussion

We have reviewed the algorithmic complexity of some of the main approaches in computational biology. These algorithms have been developed to find efficient solutions to complex and fundamental problems in biomedical research, such as sequence alignment, genome assembly, biomolecular structure determination, and many others.

Computational strategies encompassing HPC and heuristics are crucial to accelerate the solutions of complex problems in computational biology. HPC allows speeding up the computation to reasonable time scales. In particular, programs that take advantage of parallelism have been successfully applied to a number of relevant problems in computational biology (Table 3), proving that the basic computational motives of parallel computing as well as combinations of those that facilitate addressing problems that were previously considered computationally intractable. HPC is crucial in finding efficient solutions for NP-complete problems and facilitate scientific and technological advances in this domain.

Likewise, the use of heuristic algorithms (Table 2) represents a common and valid strategy that enables higher efficiency of the computation process at the cost of absolute optimality. The sub-optimal solutions found are satisfactory for specific, otherwise impractical, NP-complete problems. NP-complete problems occur in connection with computational biology challenges, for which in most cases neither available time nor calculating power is sufficient to find optimal solutions. In these cases, when searching for global optimal solutions is infeasible, approximation algorithms and heuristics are the only possible options.

The ever-growing ecosystem of bioinformatics and computational biology tools and packages responds to two main needs. First, the constant emergence of new high-throughput molecular biology technologies and data types, as well as their massive accumulation in sparse and heterogeneous reservoirs, pose unprecedented challenges and increase the complexity of the old ones [85]. Second, algorithmic solutions to these problems are in most cases, if not in all, very complex or intractable and cannot be solved in realistic computational time. Indeed, the fact that many methods are developed to address those intractable problems provides plain evidence of the need for a trade-off between approximated solutions and computational performances. Given such intrinsic complexity of biomedical problems, the computational biology and bioinformatics community is actively engaged in the implementation of improved algorithmic strategies to enhance scalability and performance.

On the other hand, the implementation details (e.g. data structures, programming languages, built-in functions) of a given algorithm can greatly impact its performance. In fact, an implementation of a given algorithm may outperform a different one in a given dataset and even exhibit unexpected outcomes when tested on others. For this reason, gold standards, reference and benchmarking datasets are critical tools to correctly assess the performance of the different algorithmic solutions for a given problem, as reviewed elsewhere [86,87]. As most of the problems in computational biology are intractable and therefore the vast majority of the solutions rely on heuristic or approximate solutions, these implementations cannot be separated from the HPC resources required for their execution. Thus, given the increasing complexity of the molecular biology problems and the unstoppable introduction of new machine learning methods, the integration of algorithms and HPC will become even more indispensable in the near future.

Computational biologists and bioinformaticians are challenged with a range of complex problems, comprising the analysis and modelling of sequences and three-dimensional structures, the examination of dynamics and evolutionary relationships of biological entities and systems, inferring and simulating molecular networks, processing medical images, documents, data streams, and many more. In this context, HPC and parallel computing are becoming essential tools for addressing intensive data tasks and achieving scalable solutions to complex problems. Therefore, it is increasingly important for users and developers to be aware of the implications of computational complexity theory for a better understanding and design of efficient algorithms in bioinformatics and computational biology.

**Funding**

This work was supported by the IBM-BSC Joint Study Agreement (JSA) on Precision medicine under the IBM-BSC Deep Learning Center Agreement and by the European Commission under the INFORE project [H2020-ICT- 825070].

**Acknowledgements**

The authors wish to acknowledge the advice and support provided by José María Fernández and Miguel Vázquez on this work.